\documentclass[fleqn,usenatbib]{mnras}

\usepackage{newtxtext,newtxmath}

\usepackage[T1]{fontenc}
\usepackage{ae,aecompl}

\usepackage{graphicx}	
\usepackage{amsmath}	
\usepackage{amssymb}	

\title[Molecular-gas properties in H$1429-0028$]{The molecular-gas properties in the gravitationally lensed merger HATLAS\,J142935.3-002836}

\author[H. Messias et al.]{Hugo Messias$^{1,2}$,\thanks{E-mail: hugo.messias@alma.cl}
	Neil Nagar$^{3}$,
	Zhi-Yu Zhang$^{4}$,
	Iv\'an Oteo$^{4,5}$,
	Simon Dye$^{6}$,\newauthor
	Eduardo Ibar$^{7}$,
	Nicholas Timmons$^{8}$,
	Paul van der Werf$^{9}$,
	Dominik Riechers$^{10}$,\newauthor
	Stephen Eales$^{11}$,
	Rob Ivison$^{4,5}$,
	Jacob Maresca$^{6}$,
	Micha{\l} J. Micha{\l}owski$^{12}$,\newauthor
	Chentao Yang$^{2}$
	\\
	$^{1}$Joint ALMA Observatory, Alonso de C\'ordova 3107, Vitacura 763-0355, Santiago, Chile \\
	$^{2}$European Southern Observatory, Alonso de C\'ordova 3107, Vitacura, Casilla 19001, Santiago de Chile, Chile \\
	$^{3}$Astronomy Department, Universidad de Concepci\'on, Barrio Universitario S/N, Concepci\'on, Chile \\
	$^{4}$European Southern Observatory, Karl-Schwarzschild-Str. 2, 85748 Garching, Germany \\
	$^{5}$Institute for Astronomy, University of Edinburgh, Royal Observatory, Blackford Hill, Edinburgh EH9 3HJ UK \\
	$^{6}$School of Physics and Astronomy, University of Nottingham, University Park, Nottingham, NG7 2RD, U.K. \\
	$^{7}$Instituto de F\'isica y Astronom\'ia, Universidad de Valpara\'iso, Avda. Gran Breta\~na 1111, Valpara\'iso, Chile \\
	$^{8}$Department of Physics and Astronomy, University of California, Irvine, CA 92697, USA \\
	$^{9}$Leiden Observatory, Leiden University, PO Box 9513, 2300 RA Leiden, The Netherlands \\
	$^{10}$Cornell University, Space Sciences Building, Ithaca, NY 14853, USA \\
	$^{11}$School of Physics and Astronomy, Cardiff University, The Parade, Cardiff CF24 3AA, UK \\
	$^{12}$Astronomical Observatory Institute, Faculty of Physics, Adam Mickiewicz University, ul.~S{\l}oneczna 36, 60-286 Pozna{\'n}, Poland \\
}

\date{Accepted XXX. Received YYY; in original form ZZZ}

\pubyear{2019}

\begin{document}
	\label{firstpage}
	\pagerange{\pageref{firstpage}--\pageref{lastpage}}
	\maketitle
	
	\begin{abstract}
		Follow-up observations of (sub-)mm-selected gravitationally-lensed systems have allowed a more detailed study of the dust-enshrouded phase of star-formation up to very early cosmic times. Here, the case of the gravitationally-lensed merger in HATLAS\,J142935.3-002836 (also known as H$1429-0028$; $z_{lens}=0.218$, $z_{bkg}=1.027$) is revisited following recent developments in the literature and new APEX observations targeting two carbon monoxide (CO) rotational transitions J$_{\rm up}$=3 and 6. We show that the line-profiles comprise three distinct velocity components, where the fainter high-velocity one is less magnified and more compact. The modelling of the observed spectral line energy distribution of CO J$_{\rm up}$=2 to 6 and [CI]~$^3P_1-^3P_0$ assumes a large velocity gradient scenario, where the analysis is based on four statistical approaches. Since the detected gas and dust emission comes exclusively from only one of the two merging components (the one oriented North-South, NS), we are only able to determine upper-limits for the companion. The molecular gas in the NS component in H$1429-0028$ is found to have a temperature of $\sim$70\,K, a volume density of $\log({\rm n [cm^{-3}]})\sim3.7$, to be expanding at $\sim$10\,km/s/pc, and amounts to ${\rm M_{H_2}=4_{-2}^{+3} \times 10^9~M_\odot}$. The CO to H$_2$ conversion factor is estimated to be $\alpha_{\rm CO}=0.4_{-0.2}^{+0.3}~$M$_\odot$/(K~km/s~pc$^2)$. The NS galaxy is expected to have a factor of $\gtrsim10\times$ more gas than its companion (${\rm M_{H_2}}\lesssim3\times10^8$~M$_\odot$). Nevertheless, the total amount of molecular gas in the system comprises only up to 15\,per cent ($1\sigma$ upper-limit) of the total (dynamical) mass.
	\end{abstract}
	
	\begin{keywords}
		Gravitational lensing: strong -- Galaxies: interactions --- Submillimeter: galaxies --- Submillimeter: ISM -- ISM: abundances
	\end{keywords}
	
	
	
	\section{Introduction}
	
	Understanding the life cycle of galaxies necessitates an observational multi-wavelength approach, not only due to the many evolution tracks a galaxy may follow \citep[e.g., finishing as an early-type galaxy, or as an ``untouched'' disc galaxy as NGC\,1277,][]{trujillo14}, but also the many physical mechanisms at play (e.g., star-formation and its quenching, heavy elements production), and the advantages and disadvantages specific to different methods of analysis.
	
	Specifically, the brightest of the dusty star-forming galaxies \citep[DSFGs;][for a review]{casey14}, also referred to as Sub-Millimetre Galaxies \citep[SMGs;][]{smail97,hughes98}, provide a strong case for the early active evolution stages of the most massive galaxies seen in the local Universe \citep[e.g.,][]{simpson14,toft14}. Despite not being representative of the whole galaxy population, SMGs may enable us to unveil the dusty origins of the local Universe's massive monsters, as seen in a Universe with less than 25 per cent its current age ($z\gtrsim2$). They are highly star-forming and already massive galaxies, quickly exhausting their large gas reservoirs, and with a source density comparable to massive galaxies in the local Universe \citep{casey14}. Whether or not the SMG phase is responsible for the build up of the bulk of the stellar populations in massive galaxies today is still a matter of active discussion, but what is clear is that this short lived phase can indeed induce a significant galaxy-growth in relatively small cosmological-time intervals ($\sim100\,$Myr).
	
	A useful characteristic of DSFGs is that, among the brightest of their kind, there are easily-selected strongly-lensed systems \citep{negrello10,vieira10,wardlow13} that can be exploited to probe this intriguing galaxy population down to fainter fluxes and higher resolution. These chance alignments are nevertheless rare \citep[less than one per deg$^{2}$][]{negrello10} and require wide field surveys. These have been possible in recent years with facilities such as the \textit{Herschel} Space Observatory (\textit{Herschel}), South Pole Telescope (SPT), or \textit{Planck}, which enabled $>100\,$deg$^2$ surveys at far-infra-red (FIR) to millimetre (mm) wavelengths. The \textit{Herschel} Astrophysical Terahertz Large Area Survey \citep[HATLAS;][]{eales10} has covered different patches of the 100--500\,$\mu$m sky amounting to 570\,deg$^2$. The Herschel Stripe 82 Survey \citep[HerS;][]{viero14} and the HerMES Large Mode Survey \citep[HeLMS;][]{oliver12} covered, respectively, 280 and 95\,deg$^2$ of the sky at 250--500\,$\mu$m. The SPT covered 2500\,deg$^2$ of the southern sky at 1--3\,mm \citep{williamson11}. Finally, the Planck mission covered the whole sky at 0.35--10\,mm \citep{planck16}.
	
	There are now numerous cases of gravitationally-lensed galaxies for exploration of the finer details of galaxy evolution at earlier times. This is done by intensive follow-up campaigns of these systems, for instance, via high-resolution multi-wavelength imaging enabling rest-frame optical to FIR reconstruction of the background source-plane \citep[e.g.,][]{negrello10,calanog14,messias14,timmons16}, or optical to mm spectroscopy to constrain the distances to lens and lensed galaxies \citep[e.g.,][]{scott11,lupu12,vieira13,strandet16,negrello17} and the gas conditions and content of the star-bursting background systems \citep[e.g.,][]{riechers11,harris12,timmons15,bothwell17,strandet17,oteo17a,yang17,canameras18,motta18}.
	
	\subsection{The HATLAS\,J142935.3-002836 case} \label{sec:h1429}
	
\begin{figure}
		\centering
		\includegraphics[width=0.5\textwidth]{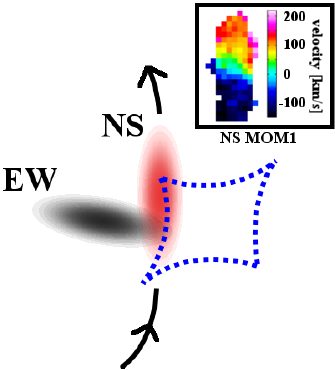}
		\caption{This toy model shows a simplified view of the background merger in H$1429-0028$. The East-West oriented galaxy (EW) dominates the rest-frame optical spectral range, while the North-South oriented one (NS) dominates the long-wavelength spectral regime. The caustic is overlaid as a dashed blue line. The currently inferred impact direction is indicated with the black arrows (see Sec.~\ref{sec:gascond}). The inset shows the velocity map (moment-1) of the NS galaxy (M14), where the colour bar extends from -150 to 150\,km/s.}
		\label{fig:toymodel}
	\end{figure}

	This manuscript reports the latest advances in comprehending the gravitationally-lensed system in HATLAS\,J142935.3-002836 \citep[also known as H$1429-0028$;][M14 henceforth]{messias14}. M14 reported the initial findings on H$1429-0028$ which appeared as the brightest  of an early set of candidates for gravitationally-lensed galaxies in HATLAS (an observed FIR flux at $160\mu{\rm m}$ of $S_{160\mu{\rm m}}\sim1.2\,$Jy). It was found that H$1429-0028$ comprises an edge-on disc galaxy at $z=0.218$ acting as lens, surrounded by an almost complete Einstein ring with a 3 to 4 knot morphology. Sub-arcsec imaging at near-IR to radio wavelengths enabled reconstruction of the background source to reveal a merging system magnified by an overall factor of $\sim10$ and comprising two distinct galaxies at $z=1.027$ (M14). One galaxy has an East-West orientation dominating the rest-frame optical spectral range (henceforth known as the EW galaxy/component). The other, appearing North-South oriented and with a compact intrinsic half-light radius $r_{1/2}=0.9\pm0.3\,$kpc, dominates the gaseous and dusty content of the system, being the main contributor to the long-wavelength spectral regime flux (henceforth known as the NS galaxy/component). In \citet{dye18}, the magnification factor of the dust component in NS was revised to be $\sim23$, further highlighting the need for proper magnification lensing analysis incorporating complex morphologies. Figure~\ref{fig:toymodel} shows a toy model to help understand the background system. Specifically the inset shows the velocity map of the NS galaxy, where it is clear that its southern region is more blue shifted than the northern one where a peak in velocity dispersion is seen (M14). A more detailed discussion can be found in M14 (specifically, Figures~1 and 8 therein).
	
	In M14, the information available at the time only allowed for the determination of the expected range of molecular gas content. As a result, APEX observations were conducted to observe extra Carbon Monoxide (CO) transitions, thus improving its spectral line energy distribution (SLED), and, together with the already observed [CI]\,$^3P_1-^3P_0$, enable us to assess the gas conditions and content in H1429. Such observations are detailed in Sec.~\ref{sec:observ}, whose results are used together with the lines reported in M14 to pursue a Large Velocity Gradient (LVG) analysis of the CO+[CI] SLED in Sec.~\ref{sec:results}. The implications of the results presented here are discussed in Sec.~\ref{sec:disc}. Finally, Sec.~\ref{sec:conc} presents the concluding remarks.
	
	\vspace{5mm}
	Throughout this paper, the following $\Lambda$CDM cosmology is adopted: H$_0$ = 70\,km\,s$^{-1}$\,Mpc$^{-1}$, $\Omega_{\rm M}=0.3$, $\Omega_\Lambda=0.7$
	. The Cosmic Microwave Background (CMB) is assumed to expand adiabatically \citep[T$_z = {\rm T}_0\times(1+z)$, with T$_0=2.726$;][]{muller13}. When stated, the gas mass estimates are corrected by a factor of 1.36 to account for chemical elements heavier than Hydrogen, assuming the latter comprises $73-74$ per cent of the total baryonic matter mass \citep{croswell96,carroll06}.
	
	\section{Observations} \label{sec:observ}
	
	Prior to this work, only CO\,J$_{\rm up}$=[2,4,5], [CI]\,$^3P_1-^3P_0$, and CS\,(10-9) had been observed toward H$1429-0028$. The limited number of spectral features, precluded a reliable analysis of the molecular gas content in the system. It was not clear if a two-component CO SLED existed and, given the redshifts of the lens (z=0.218) and sources (z=1.027), there was the possibility that the background CO\,(5-4) emission measured by ZSPEC at APEX was contaminated by foreground CO\,(3-2) emission. This work makes use of the previously detected lines, in addition to recent observations of CO\,J$_{\rm up}$=[3,6] transitions to remove the ambiguities mentioned. The new observations are described in the following sub-sections.
	
	\subsection{APEX}
	
	The instrument suite available at the Atacama Pathfinder EXperiment (APEX) has contributed strongly toward the understanding of H$1429-0028$, giving the initial redshift determination of the system and detecting four CO transitions. The Z-SPEC observations targeting the CO transitions J=4-3 and 5-4 were already described in detail in M14. Here, we proceed to describe the more recent SEPIA Band\,5 and SHeFI APEX2 observations.
	
	\subsubsection{SEPIA Band\,5}
	
	The CO\,(3-2) and CS\,(7-6) transitions were targeted with APEX/SEPIA-Band\,5 between 25$^{\rm th}$\,May and 5$^{\rm th}$\,June 2016 (097.A-0995, P.I. Messias). The lower-side (signal) band was tuned to 170\,GHz to cover the two lines of interest, putting the upper-side (image) band at 182\,GHz. The observations were conducted under {\sc pwv}$\sim$3\,mm (ranging from 2.6 to 5.1\,mm between 25$^{\rm th}$\,May and 1$^{\rm st}$\,June, and 1.3 to 1.7\,mm on 5$^{\rm th}$\,June) using either Jupiter or Mars as calibrators. A standard observing strategy was adopted (wobbler-switching with a 50'' amplitude, at a frequency of 0.5\,Hz). The total time on source was 5.1\,h. The data was reduced with {\sc class} (a {\sc gildas}\footnote{http://www.iram.fr/IRAMFR/GILDAS} task). The {\sc rms} uncertainty on the final co-added spectrum is 0.3\,mK (12\,mJy) at a spectral resolution of 100\,km/s. The adopted Kelvin to Jansky conversion factor is 38.4\,Jy/K \citep{belitsky18}. 
	
	\subsubsection{SHeFI APEX2}
	
	The CO\,(6-5) transition was targeted with APEX/SHeFI-APEX2 on 5$^{\rm th}$\,June and between 29$^{\rm th}$\,July and 4$^{\rm th}$\,August 2016 (097.A-0995, P.I. Messias). The lower-side (signal) band was tuned to 339.8\,GHz to cover the line of interest and CS\,(14-13), putting the upper-side (image) band at 351.8\,GHz. The observations were conducted under {\sc pwv}=0.55--1.4\,mm, using IRAS$15194-5115$, SW Vir, and SgrB2(N) as calibrators. A standard observing strategy was adopted (wobbler-switching with a 50'' amplitude, at a frequency of 0.5\,Hz). The total time on source was 4\,h. The data was reduced with {\sc class} (a {\sc gildas} task). The {\sc rms} uncertainty on the final co-added spectrum is 0.6\,mK (24\,mJy) at a spectral resolution of 100\,km/s. The adopted Kelvin to Jansky conversion factor is 40.8\,Jy/K following the procedure reported in the APEX website\footnote{http://www.apex-telescope.org/telescope/efficiency/index.php}.

	\section{Results} \label{sec:results}
	
	\subsection{Properties of targeted lines} \label{sec:lines}
	
\begin{figure}
		\centering
		\includegraphics[width=0.5\textwidth]{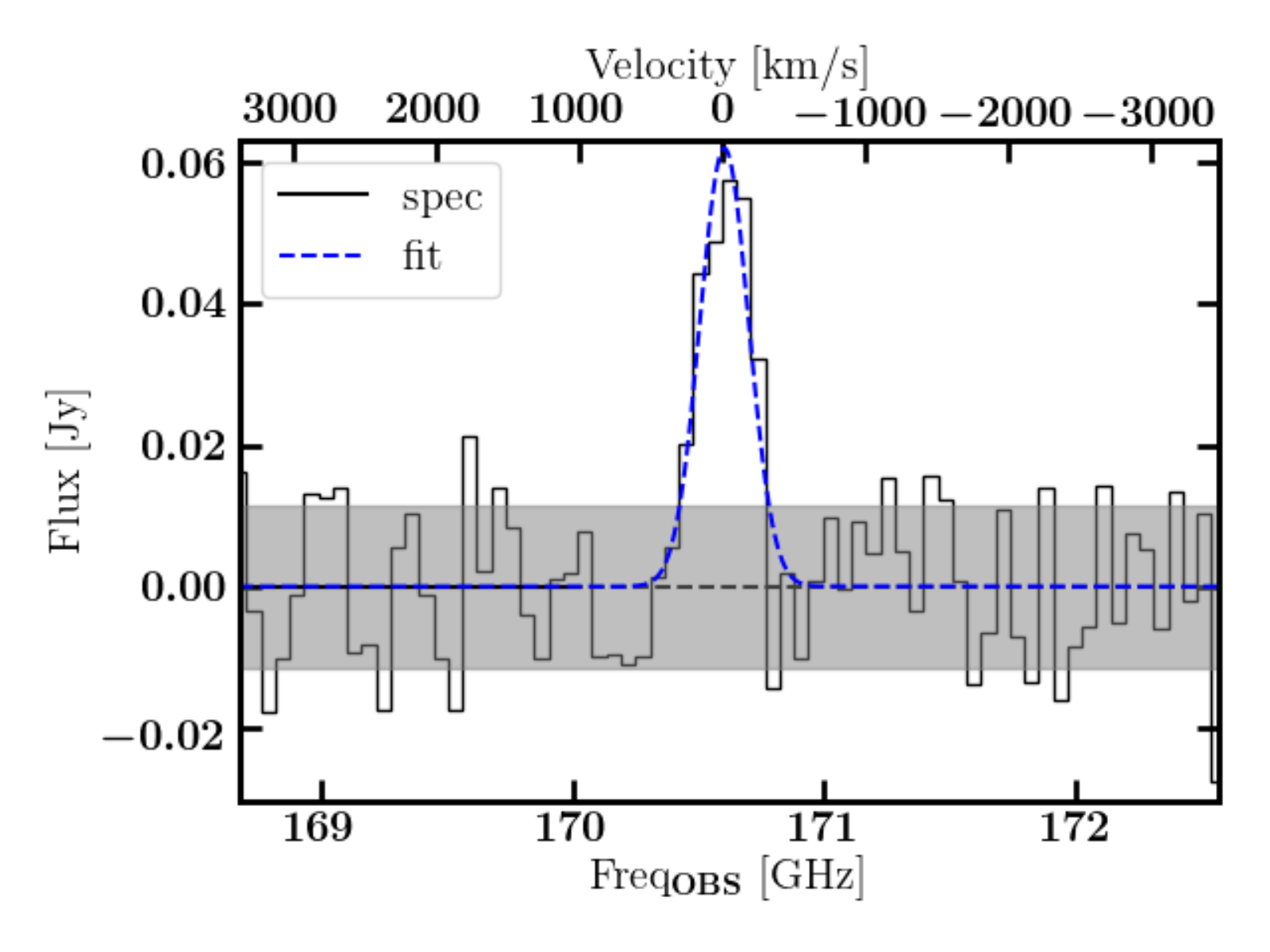}
		\caption{The SEPIA-Band\,5 spectrum targeting CO\,(3-2) transition. The observed spectrum is shown as the solid black histogram. The Gaussian fit to the line-emission is shown as a dashed blue line. The velocity frame used in the upper axis is set to $z=1.027$. The shaded grey region shows the $\pm1\sigma$ uncertainty.}
		\label{fig:co32}
	\end{figure}

\begin{figure}
		\centering
		\includegraphics[width=0.5\textwidth]{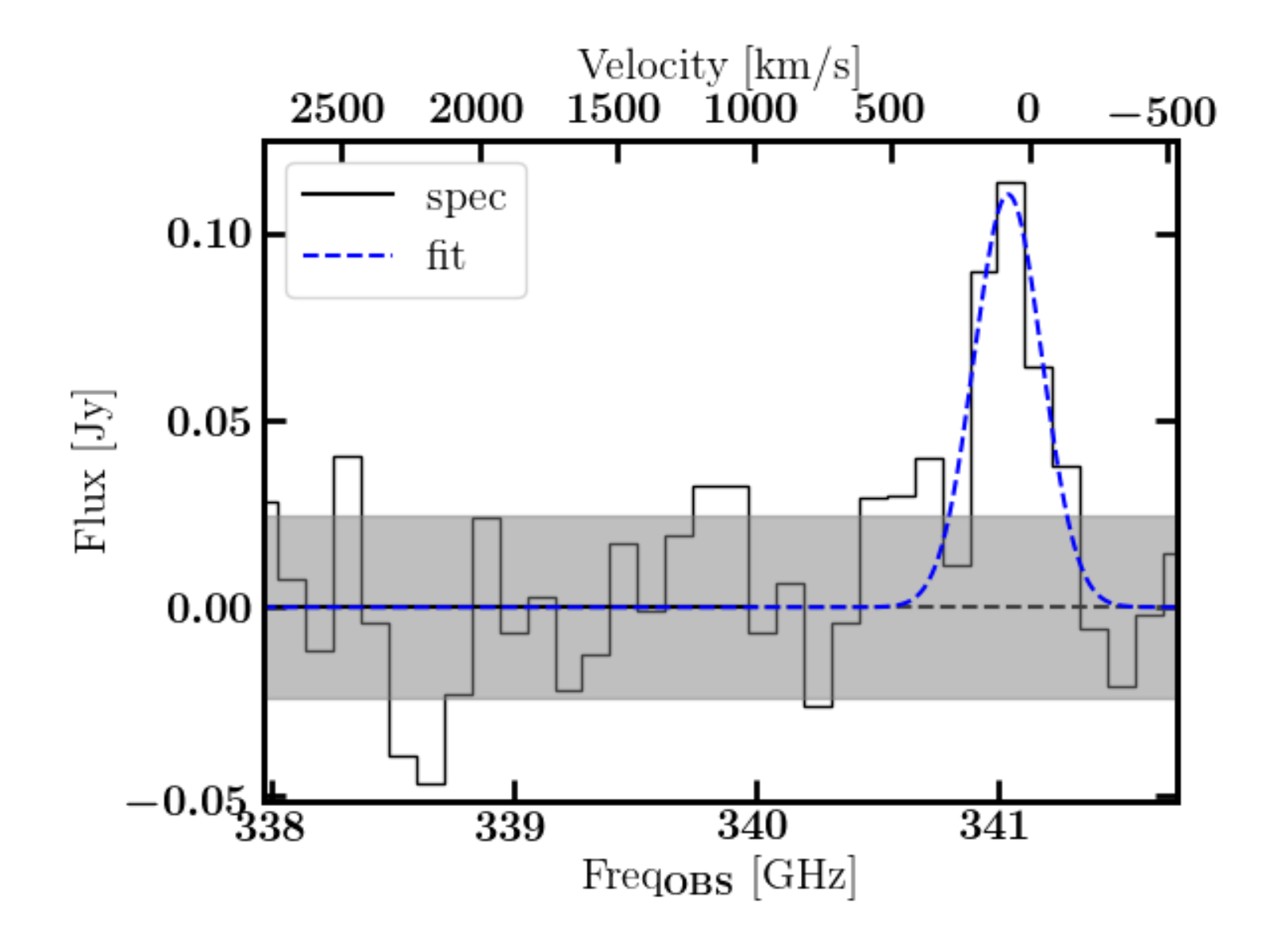}
		\caption{The same as in Figure~\ref{fig:co32}, but for the SHeFI-APEX2 spectrum targeting CO\,(6-5) transition.}
		\label{fig:co65}
	\end{figure}

	Altogether, three chemical species (C, CO, CS) and nine transitions have been targeted thus far (M14 and this work). Of these, only CS\,(7-6) and (14-13), observed by APEX, were not detected, as expected by the depths of these observations. Also, since CO\,(5-4) has not been spectrally resolved thus far, and because of the possibility of foreground line contamination (Section~\ref{sec:h1429}), this transition is left out of the analysis. Figures~\ref{fig:co32} and \ref{fig:co65} show the newly targeted CO transitions J=3-2 and 6-5 as observed by SEPIA-Band\,5 and SHeFI-APEX2, respectively.
	
\begin{figure*}
		\centering
		\includegraphics[width=1\textwidth]{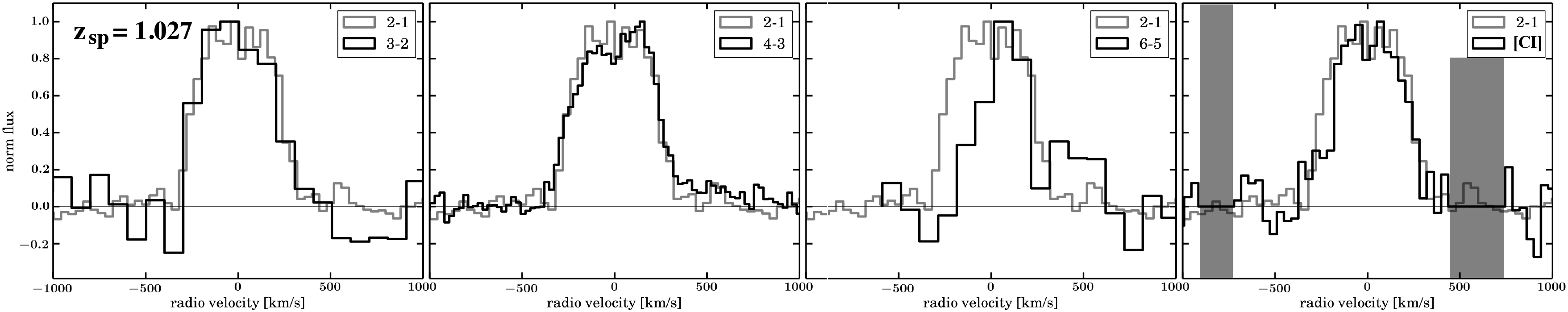}
		\caption{Line-profile comparison between the spectrally-resolved CO transitions detected to date. In each panel, CO~(2-1) is used as reference to ease comparison with CO\,J=3-2 (left panel), 4-3 (middle left), 6-5 (middle right), and [CI]\,$^3P_1-^3P_0$. The line-fluxes are normalised to the line peaks. The velocity frame is set to $z_{\rm sp}=1.027$. The J=4-3 is displayed at a $\sim$25\,km/s spectral resolution, J=2-1 and [CI] at $\sim$40\,km/s, while J=3-2 and 6-5 at $\sim$100\,km/s. This figure shows that there are up to three detected components comprising the line profile in this system: two central main ones blue- and red-shifted, and another one redshifted to $\sim500\,$km/s. The latter can not be assessed in the [CI] transition since it has been flagged due to sky-line contamination (shaded regions).}
		\label{fig:co21all}
	\end{figure*}

	Figure~\ref{fig:co21all} compares the line profiles of the CO transitions detected to date. As already mentioned in M14, the J=2-1 and 4-3 transitions show a double-peak or plateau line-profile. These new APEX observations actually follow this scenario, where the blue-shifted component is stronger at low-J transitions (and almost absent at J=6-5), while the redshifted one is stronger at high-J transitions. This could be a consequence of the red-component being more excited than the blue one as a result of the merger. Henceforth, these blue and red velocity components are referred to as components {\sc I} and {\sc II}, respectively.
	
	Interestingly, the additional flux at even higher velocities ($\sim500\,$km/s, also mentioned in M14; component {\sc III} henceforth) matches the excess flux seen at $\sim500\,$km/s in the J=6-5 spectrum. This could be evidence for a shock-induced component, like an outflow. Due to this result, we revisited the HST grism data assessing any possible emission from the arc-like feature external to the Einstein-ring (best seen in HST/WFC3-\textit{F110W} imaging), in search of evidence of a velocity match between that component and the redshifted one reported here. However, due to the depth of the data and to neighbouring source contamination, no emission is observed. The location and morphology of this feature is addressed later in Section~\ref{sec:diffmag}.
	
	\subsection{Spectral Line decomposition} \label{sec:linedec}
	
\begin{figure}
		\centering
		\includegraphics[width=\columnwidth]{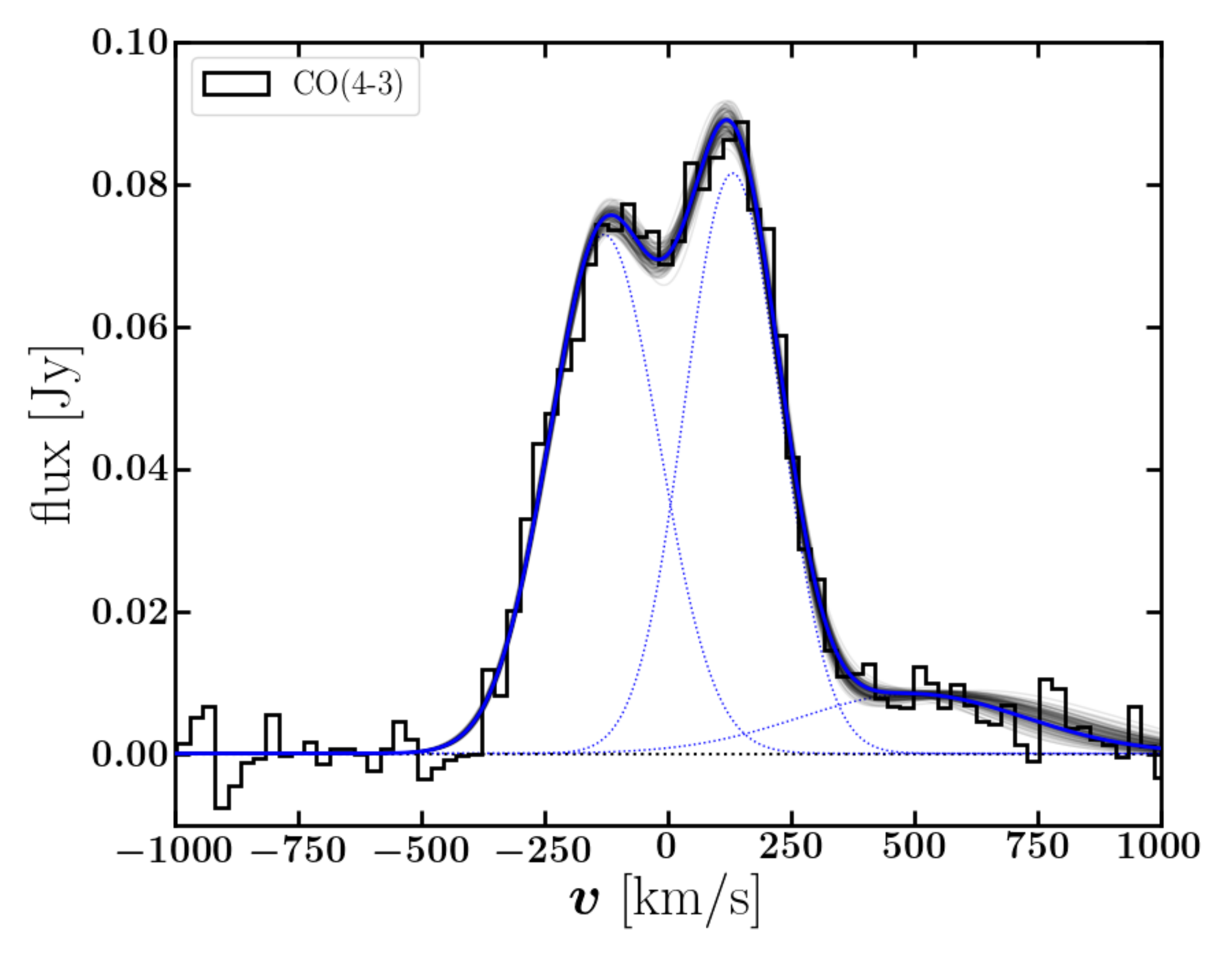}
		\caption{Spectral decomposition of the CO\,(4-3) transition. The solid black histogram shows the observed spectrum. The solid blue line represents the 50$^{\rm th}$ percentile spectral fit, with each of the three components shown individually as dotted blue lines. The grey thin solid lines are 100 randomly chosen MCMC samples. This analysis served as a training step before fitting the remainder spectrally resolved transitions (Fig.~\ref{fig:specdec}). The error level per-channel relative to peak is $\sim$1.2\,per cent.}
		\label{fig:co43train}
	\end{figure}

	The spectral decomposition of the line-profiles into the three possible spectral components mentioned in the previous section was made in a train-and-fit approach. First, a three-Gaussian fit to the spatially-integrated CO\,(4-3) spectrum was pursued as a training step by making use of the Python implementation of the affine-invariant ensemble sampler for Markov Chain Monte Carlo (MCMC) proposed by \citet{goodman10}, {\sc emcee} \citep{foremanmackey13}. This transition was chosen being the one with the highest signal-to-noise ratio allowing for a better line-profile characterisation, and the result can be seen in Figure~\ref{fig:co43train}. The training-derived values of the three centroid-velocities were kept fixed while fitting the remainder of the spectrally resolved CO transition spectra. The training-derived Gaussian FWHM values of each component were considered as first guesses. Based on the line-profile discussion in Section~\ref{sec:lines} (see also Figure~\ref{fig:co21all}), the flux contribution of each velocity component to each spectrum total flux of a given $J \rightarrow J-1$ rotational transition ($\mathcal{C}^i_{\rm J,J-1}={\rm S^i/S^{\rm TOT}}$, where ${\rm S^i}$ is the velocity-integrated flux of a given velocity component $i\equiv [{\sc I,~II,~III}]$, and ${\rm S^{\rm TOT}}$ the total velocity-integrated flux) had to comply with the following priors:
	$\mathcal{C}^{\rm I}_{21}>\mathcal{C}^{\rm I}_{32}>\mathcal{C}^{\rm I}_{43}>\mathcal{C}^{\rm I}_{65}$, 
	$\mathcal{C}^{\rm II,III}_{21}<\mathcal{C}^{\rm II,III}_{32}<\mathcal{C}^{\rm II,III}_{43}<\mathcal{C}^{\rm II,III}_{65}$. Note that the $\mathcal{C}_{43}$ values are those corresponding to the 50$^{\rm th}$ percentile reported from the training step. While fitting the [CI]~$^3P_1-^3P_0$ profile, both the component centroid velocities and FWHMs were fixed to the 50$^{\rm th}$ percentiles reported from the training step. This was done since, although the line-profile shows a plateau, it also reveals a sharp break at $\sim$180\,km/s and no extra assumptions on the relative component contribution can be adopted like in the other CO transitions\footnote{We revisited the data-set to improve the self-calibration step (only in phase) and retrieved the spectrum from the image itself in a common region to CO\,(4-3) spectrum extraction, not from the model component map as in M14. This resulted in the differences between the [CI]~$^3P_1-^3P_0$ spectrum shown here and in M14.}. Due to sky-line contamination, component III is not observable in the [CI] transition.
	
\begin{figure*}
		\centering
		\includegraphics[width=\textwidth]{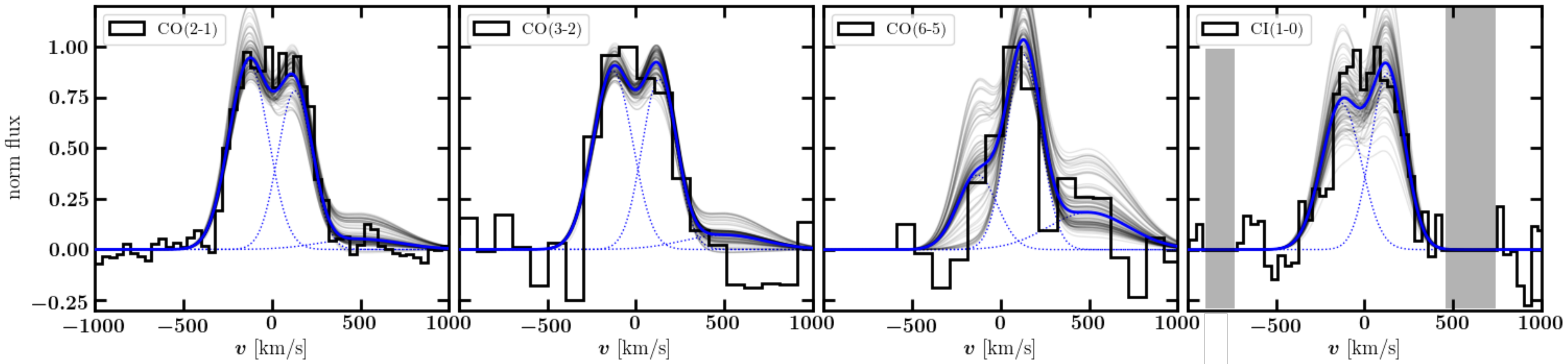}
		\caption{Spectral decomposition of the other spectrally resolved CO and [CI] transitions. The solid black histograms show the observed spectra normalised to maximum flux for comparison purposes. The solid blue line represents the 50$^{\rm th}$ percentile spectral fit, with each of the three components shown individually as dotted blue lines. The grey thin solid lines are 100 randomly chosen MCMC samples. The training step was done by fitting the CO~(4-3) transition (Fig.~\ref{fig:co43train}). The central velocities of each spectral component were kept fixed. The error level per-channel relative to peak is about: 2.6\,per cent (left panel), 20\,per cent (middle left), 22\,per cent (middle right), 3.2\,per cent (right). Again, note that the shaded regions in the [CI] panel show the velocity ranges contaminated by atmospheric-line.}
		\label{fig:specdec}
	\end{figure*}

	Figure~\ref{fig:specdec} shows the results of this fitting approach, while Table~\ref{tab:lines} reports on the velocity-integrated fluxes of each component at each CO and [CI] transitions. The {\sc I}, {\sc II}, and {\sc III} refer to each component ordered by increasing centroid velocity ($v_c^{I}=-130_{-6}^{+6}$\,km/s, $v_c^{II}=131_{-5}^{+5}$\,km/s, and $v_c^{III}=500_{-30}^{+20}$\,km/s as measured in CO\,J:4-3). Note that components {\sc III} and {\sc I} in CO~(2-1) and (6-5), respectively, are allowed to be zero, i.e., not to be present, yet the analysis shows that these components are detected at the 3.2 and 1.4$\sigma$ levels ($0.9_{-0.3}$\,Jy.km/s and $11_{-8}$\,Jy.km/s), respectively.
	
	\begin{table*}
		\caption{Properties of targeted lines toward H$1429-0028$ used in the analysis.}              
		\label{tab:lines}      
		\centering                                      
		\begin{tabular}{ccrrrrrrrc}          
			\hline\hline                        
			Species & Transition & $S \Delta v^{a}_{obs}$ & $S \Delta v^{\rm I}$ & FWHM$^{\rm I}$ & $S \Delta v^{\rm II}$ & FWHM$^{\rm II}$ & $S \Delta v^{\rm III}$ & FWHM$^{\rm III}$ & Facility \\    
			& & [Jy~km/s] & [Jy~km/s] & [km/s] & [Jy~km/s] & [km/s] & [Jy~km/s] & [km/s] & \\
			\hline                                   
			CO & 2-1 & $14.4\pm0.2$ & $7.7_{-0.7}^{+0.7}$ & $263_{-2}^{+5}$ & $5.6_{-0.8}^{+0.6}$ & $228_{-7}^{+1}$ & $0.9_{-0.3}^{+2}$ & $536.2_{-4}^{+0.1}$ & ALMA \\
			& 3-2 & $22\pm5$ & $14.3_{-0.7}^{+1}$ & $262.7_{-0.2}^{+7}$ & $11.7_{-1}^{+0.9}$ & $227.6_{-0.9}^{+2}$ & $2_{-1}^{+2}$ & $536_{-2}^{+1}$ & APEX \\
			& 4-3 & $44.8\pm0.4$ & $20.4_{-0.9}^{+0.9}$ & $263_{-7}^{+8}$ & $20_{-1}^{+1}$ & $228_{-8}^{+9}$ & $4.8_{-0.4}^{+0.5}$ & $540_{-40}^{+40}$ & ALMA \\
			& 6-5 & $36\pm10$ & $12_{-8}^{+5}$ & $263_{-3}^{+4}$ & $26_{-2}^{+4}$ & $227.6_{-6}^{+0.9}$ & $12_{-5}^{+11}$ & $536.3_{-0.2}^{+5}$ & APEX \\
			{[}CI{]} & $^3P_1-^3P_0$ & $11.9\pm0.2$ & $5.7_{-0.9}^{+1}$ & (263)$^{b}$ & $6_{-1}^{+1}$ & (228)$^{b}$ & \dots & \dots & ALMA \\
			\hline                                             
			\multicolumn{10}{l}{\footnotesize Note --- The central velocities of each component as measured in CO\,(4-3) are:  $v_c^{I}=-130_{-6}^{+6}$, $v_c^{II}=131_{-5}^{+5}$, and $v_c^{III}=500_{-30}^{+20}$}\,km/s\\
			\multicolumn{10}{l}{\footnotesize $^{a}$ Values are those retrieved by integrating the observed spectra from -500\,km/s to 1000\,km/s in CO transitions and to 500\,km/s}\\
			\multicolumn{10}{l}{in the [CI] transition.}\\
			\multicolumn{10}{l}{\footnotesize $^{b}$ The FWHM values for each of the two components in [CI]~$^3P_1-^3P_0$ were fixed to those measured in CO~(4-3).}
		\end{tabular}
	\end{table*}
	
\begin{figure*}
		\centering
		\includegraphics[width=\textwidth]{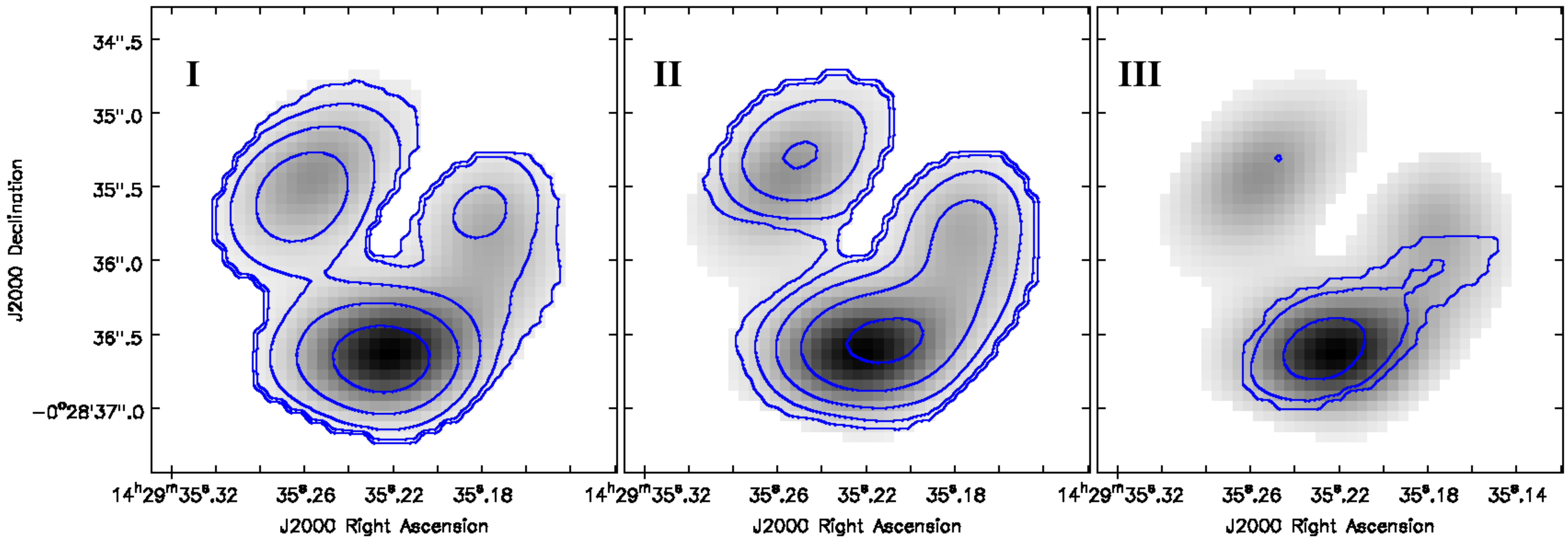}
		\caption{A spectral-spatial decomposition of the three spectral components comprising the CO\,(4-3) emission. The grey-scale image represents the total velocity-integrated flux map (moment-0). The blue contours in the left, middle, and right panels reveal the emission of, respectively, the components with centroids at around $-130$ (I), 130 (II), and 500\,km/s (III) away from the systemic velocity. The contour levels are 3, 6, 12, 24, and 48$\sigma$.}
		\label{fig:decompimg}
	\end{figure*}

	The spectral-spatial decomposition of the three components was also pursued on a pixel-by-pixel basis (i.e., not in the visibility plane). Again, a three-Gaussian fit to the spatially-integrated CO(4-3) spectrum was used as a training step. The training-derived values of the centroid-velocity and Gaussian FWHM of each component were used as reference for each spectrum fit in each cube-pixel, while the relative amplitudes were computed independently for each cube-pixel. Figure~\ref{fig:decompimg} shows the result of this spatial-spectral decomposition via the velocity-integrated flux maps (moments-0) of each studied component. The spatial distributions are distinct between the three, hence coming from different regions and this may result in distinct magnifications.
	
	\subsection{Differential magnification} \label{sec:diffmag}
	
	Given the extension and multiplicity of the background system, one may wonder whether differential magnification may occur to some degree. In table~3 of M14, the lensing modelling at the time showed a range in the magnification factor between 5 and 11 depending on the spectral band. Since then, new observations have been obtained at much finer spatial-resolution \citep[0.12~arcsec;][]{dye18} tracing the dust emission at rest-frame 696\,GHz (430\,$\mu$m). These allowed an improvement of the lens model, which resulted in a revised magnification factor of 24$\pm$1, twice that reported in M14 (10.8$\pm$0.7). Even though the latter was based on slightly lower-frequency data (rest-frame 474\,GHz or 632\,$\mu$m), both are tracing cold dust via the Rayleigh-Jeans tail of the spectrum, hence the difference between the two results from the improvement in the lensing model.
	
\begin{figure*}
		\centering
		\includegraphics[width=\textwidth]{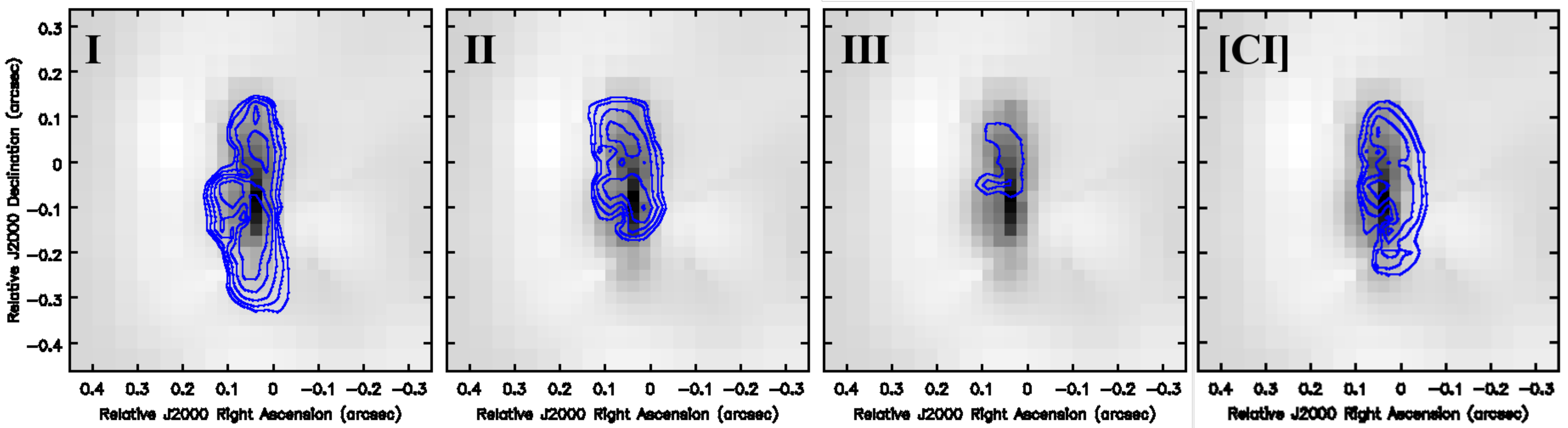}
		\caption{The three panels show the spectral-spatial decomposition in the $0.8\times0.8\,$arcsec source-plane of the three spectral components comprising the CO\,(4-3) emission. The grey-scale image represents the total velocity-integrated flux map (moment-0). The blue contours in the left, middle-left, and middle-right panels reveal the CO\,(4-3) emission of, respectively, the components with centroids at $-130_{-6}^{+6}$\,km/s ({\sc I}), $+131_{-5}^{+5}$\,km/s ({\sc II}), and $+500_{-30}^{+20}$\,km/s ({\sc III}) away of the systemic velocity. The blue contours in the right panel reveal the [CI]\,$^3P_1-^3P_0$ velocity-integrate flux map (components {\sc I} and {\sc II} were considered together). The contour levels are set at the 3, 4.2, 6, 8.5, and 12$\sigma$.}
		\label{fig:decomp}
	\end{figure*}

	We have thus revised the source reconstruction of the CO\,(4-3) emission. Specifically, each moment-0 map of the spectral components (Figure~\ref{fig:decompimg}) was reconstructed separately using the same lens model from \citet{dye18}, and AutoLens \citep{nightingale18} was used as an independent check of the lens parameters. The best-fit result is seen in Figure~\ref{fig:decomp}, where the source areas (A) by counting the number of pixels in the source-plane above $3\sigma$ are found to be A$^{\sc I}=0.062\,$arcsec$^2$ (4.1\,kpc$^2$), A$^{\sc II}=0.039\,$arcsec$^2$ (2.6\,kpc$^2$), and A$^{\sc III}=0.011\,$arcsec$^2$ (0.72\,kpc$^2$). This new analysis also implies that the current estimate for the dynamical mass and half-light radius of the NS component in the merger is M$_{\rm DYN}=5_{-2}^{+3}\times10^{10} \,$M$_\odot$ and $r_{1/2}=0.7_{-0.3}^{+0.5}\,$kpc, which are in agreement with the reported values in M14, where the ``isotropic virial estimator'' was also assumed: ${\rm M_{dyn}}=2.8\times10^5 (\Delta v_{\rm FWHM})^2 r_{1/2}$, where $[{\rm M_{dyn}}]=M_\odot$, $\Delta v_{\rm FWHM}$ is the CO(4-3) FWHM, and $r_{1/2}$ is the half-light radius in units of kpc.
	
\begin{figure}
		\centering
		\includegraphics[width=0.5\textwidth]{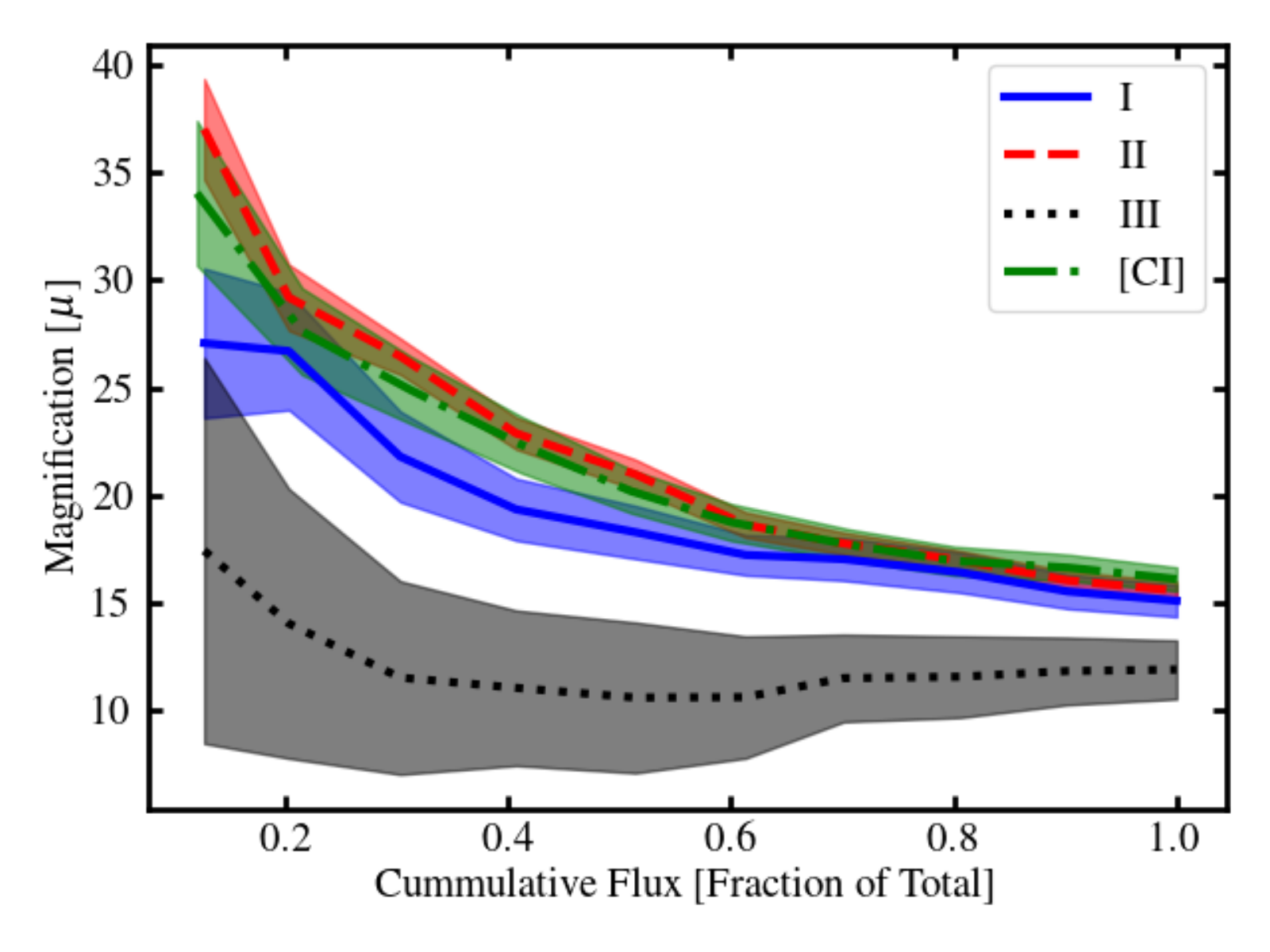}
		\caption{Assessing the differential magnification in H$1429-0028$. The cumulative source flux (indicated as fraction of total flux) is estimated by gradually summing source-plane pixels in decreasing flux order. While components {\sc I} and {\sc II} in CO(4-3) are magnified by a comparable amount ($\mu\sim15$), component {\sc III} is less magnified ($\mu\sim12$). The profile of [CI]\,$^3P_1-^3P_0$ for the combined emission from components {\sc I} and {\sc II} is also displayed in green, showing no significant difference with respect to CO(4-3).}
		\label{fig:magprof}
	\end{figure}

	The magnification profiles for each component are displayed in Figure~\ref{fig:magprof}. These show how the magnification evolves with cummulative source flux (indicated as fraction of total flux), which is estimated by gradually summing source-plane pixels in decreasing flux order. It is evident that component {\sc iii} is consistently less magnified than the other two components, which have indistinguishable total magnifications at flux fractions above $\sim0.6$ within errors. The total magnifications are the following: $\mu^{\sc I}=15.1\pm0.7$, $\mu^{\sc II}=15.6\pm0.4$, and $\mu^{\sc III}=12\pm1$. 
	
	Finally, it is worth assessing the differential magnification between each emission line. The other transition resolved into the ring and knot morphology observed in CO(4-3) is [CI]\,$^3P_1-^3P_0$. The same model from \citet{dye18} was applied to the velocity-integrated flux map of [CI]. This time, for the purpose of maximising signal-to-noise ratio, and since the CO(4-3) analysis pointed to no differential magnification between the velocity components {\sc I} and {\sc II}, both of these were analysed together (i.e., no spectral deblending was considered). The spatial distribution of the emission at $>3\sigma$ is displayed in Figure~\ref{fig:decomp}, comprising an area of A$_{\sc [CI]}=0.046\,$arcsec$^2$ (3.0\,kpc$^2$). The magnification profile is displayed in Figure~\ref{fig:magprof}, where it is seen that it follows those observed in CO(4-3) and its total magnification factor is 16.1$\pm$0.6.
	
	As a result, based on the available data, we assume below that the analysis is not affected by differential magnification between velocity-components nor between spectral lines. Specifically, we assume that the velocity components {\sc I} and {\sc II} are equally magnified as well as the CO transitions between themselves and with respect to [CI].
	
	\subsection{Large Velocity Gradient analysis} \label{sec:lvg}
	
	In Section~\ref{sec:diffmag}, we make the case that, within the uncertainties, the differences between the two main spectral components {\sc I} and {\sc II} are not affected by differential magnification, but rather from different excitation levels \citep{daddi15,yang17,canameras18}. Also, component {\sc III} is significantly less magnified and its [CI] emission is unconstrained due to sky-line contamination, preventing a proper analysis of its physical conditions. However, given its brightness, it is expected to account for a relatively minor fraction of the molecular gas budget in the system. Nevertheless, follow-up observations are of interest since this feature will likely help understand the current stage of evolution of the merger (e.g., if it is confirmed to be an outflow). As a result, this section will focus on the separate analysis of components {\sc I} and {\sc II} alone. We note that it does not consider CO\,(5-4) since it is not spectrally resolved. In the future, high spatial-resolution imaging will enable proper multi-J spectral and spatial decomposition of the CO emission as it is shown for CO\,(4-3) in Section~\ref{sec:diffmag}.
	
	As already mentioned, many transitions were detected thus far toward H1429-0028: CO\,(J$_{\rm up}=2$, 3, 4, 5, 6), [CI]\,($^3P_1-^3P_0$), and CS\,(10-9). The former two species are commonly used to indirectly derive the molecular gas content. While CO is brighter, thus easier to detect, the fine structure system of [CI] is characterised by a simple three-level system easily excited by particle collisions, and is believed to be widespread in giant molecular clouds \citep[][and references therein]{tomassetti14}. The CS molecule is regarded as a high-density gas tracer, hence it could be used as well, but this retrieved poor results (e.g., the predicted CO SLED overestimated the high-J transitions' fluxes). This may be a result of CS being detected in a very small region in the system, while CO and C are more extended, which could then result in differential magnification (Figure~\ref{fig:magprof}), further enhancing the difference. 
	
	We adopt the large velocity gradient (LVG) formalism to interpret the CO\,(J$_{\rm up}=2$, 3, 4, 6) and [CI] SLED by making use of \textit{myRadex}\footnote{https://github.com/fjdu/myRadex}. An escape probability of $\beta=\frac{1-e^{-\tau}}{\tau}$ is assumed in an expanding spherical geometry. The CMB temperature is 5.53\,K at $z=1.027$. The CO line width was set to 262 and 228\,km/s for components {\sc I} and {\sc II}, respectively, in accordance to the values in Tab.\ref{tab:lines}. The CO and C abundances relative to H$_2$ (small $x_{\rm CO}$ and $x_{\rm C}$) are both assumed to be $\sim10^{-4}$, which already implies that a significant fraction of Carbon is locked into CO \citep{walter11}, i.e., the CO emission mostly comes from dense molecular regions \citep[e.g.,][]{wolfire10,narayanan12}. The uncertainty in these abundances is further explored in Section~\ref{sec:cih2}.
	
	We have adopted different approaches to estimate the gas temperature (T~[K]), the molecular number density ($n_{\rm H_2}$~[cm$^{-3}$]), velocity gradient ($dv$~[m/s/pc]), column density (N$_{\rm H_2}$~[cm$^{-2}$])\footnote{Note that the following is assumed: N$_{\rm H_2}=n_{\rm H_2}\times\frac{\rm line_{FWHM}}{dv}\times x_{C,CO}$}, and the molecular-gas mass (${\rm M_{H_2}}$~[M$_\odot$]), while considering CO and [CI] together. The best-fit value was obtained by finding the minimum $\chi^2$-value in a grid of line-intensity values considering conditions ranging $10<{\rm T~[K]}<1000$, $10^2<n_{\rm H_2[cm^{-3}]}<10^6$, and $10^2<dv{\rm~[m/s/pc]}<10^6$, with $\log$-steps of 0.1\,dex. Maximum Likelihood \citep[ML; e.g.,][]{zhang14}, Bootstrapping, and MCMC approaches also adopted this same grid and provided the 16$^{\rm th}$, 50$^{\rm th}$, and 84$^{\rm th}$ percentiles. While Bootstrapping, a total of 1000 iterations were adopted where, in each one, the observed line-flux values were randomly defined to be around the real observed value following a Gaussian distribution with a standard deviation equal to the estimated flux error. This was adopted in all the above approaches, together with a 5 and 10\,per cent error in flux added in quadrature to the instrumental error to account for the uncertainty in the absolute flux scaling in CO\,(2-1) and the remaining transitions, respectively.
	
	This analysis adopted priors in order to better constrain the results. The H$_2$-mass should not be larger than the dynamical-mass (M${\rm _{H_2}}<7.8\times10^{10}\,{\rm M_\odot}$, $+1\sigma$ of the dynamical-mass estimate). It is also assumed that the gas is in virial equilibrium or super-virialised, i.e., the velocity gradient is equal or greater than that calculated from the virial equilibrium ($K_{vir}=(\rm{d}v/\rm{d}r)_{\rm LVG}~/~(\rm{d}v/\rm{d}r)_{\rm vir}\geq1$) as expected from normal to star-burst galaxies \citep[e.g.,][]{papadopoulos99,zhang14}.
	
\begin{figure*}
		\centering
		\includegraphics[width=\textwidth]{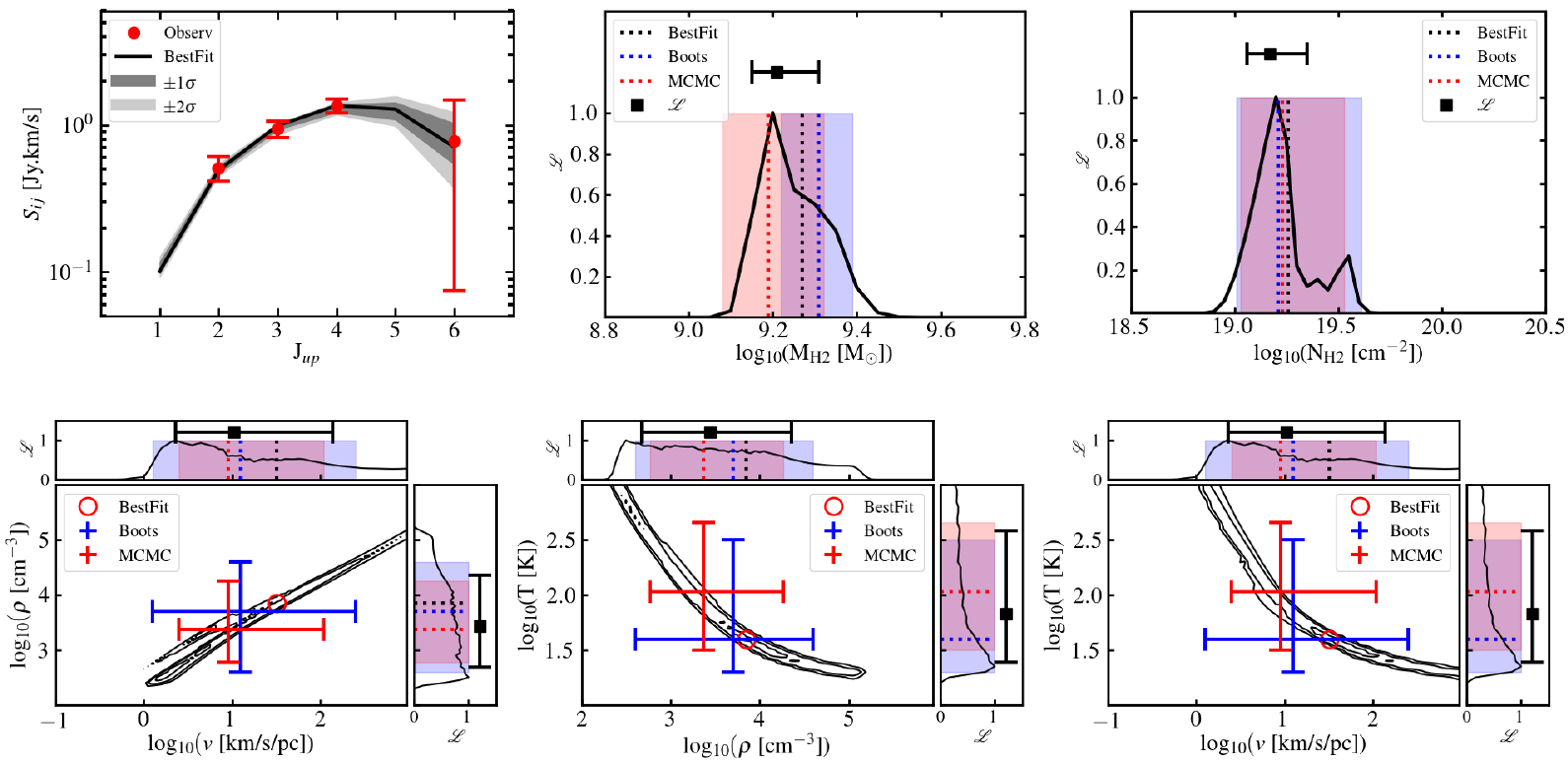}
		\caption{Modelling H1429-0028's CO and [CI] SLED with \textit{myRadex} for component\,{\sc I}. [\textit{Top Left}] The solid black line shows the best fit to the CO SLED, while the dark and light grey shaded regions show the 16$^{\rm th}$ to 84$^{\rm th}$ and 2$^{\rm th}$ to 98$^{\rm th}$ percentile ranges of the model predicted-fluxes resulting from the maximum likelihood analysis. [\textit{Top Middle and Right}] These panels show the likelihood distributions of the H$_2$ mass and column density and their 15.9$^{\rm th}$, 50$^{\rm th}$, 84.1$^{\rm th}$ percentile values as error bars. The best-fit value is marked with a dotted line, while the blue and red dotted lines and transparent regions show the 15.9$^{\rm th}$, 50$^{\rm th}$, 84.1$^{\rm th}$ percentile values for, respectively the Bootstrapping and MCMC analysis. [\textit{Bottom}] These three panels show the probability surface density of the molecular gas density ($n_{\rm H_2}$), velocity gradient ($v$), and gas temperature (T), when comparing each with the other two. The top and side plots follow the same colour and pattern coding as the middle and right top panels.}
		\label{fig:lvgc1}
	\end{figure*}

\begin{figure*}
		\centering
		\includegraphics[width=\textwidth]{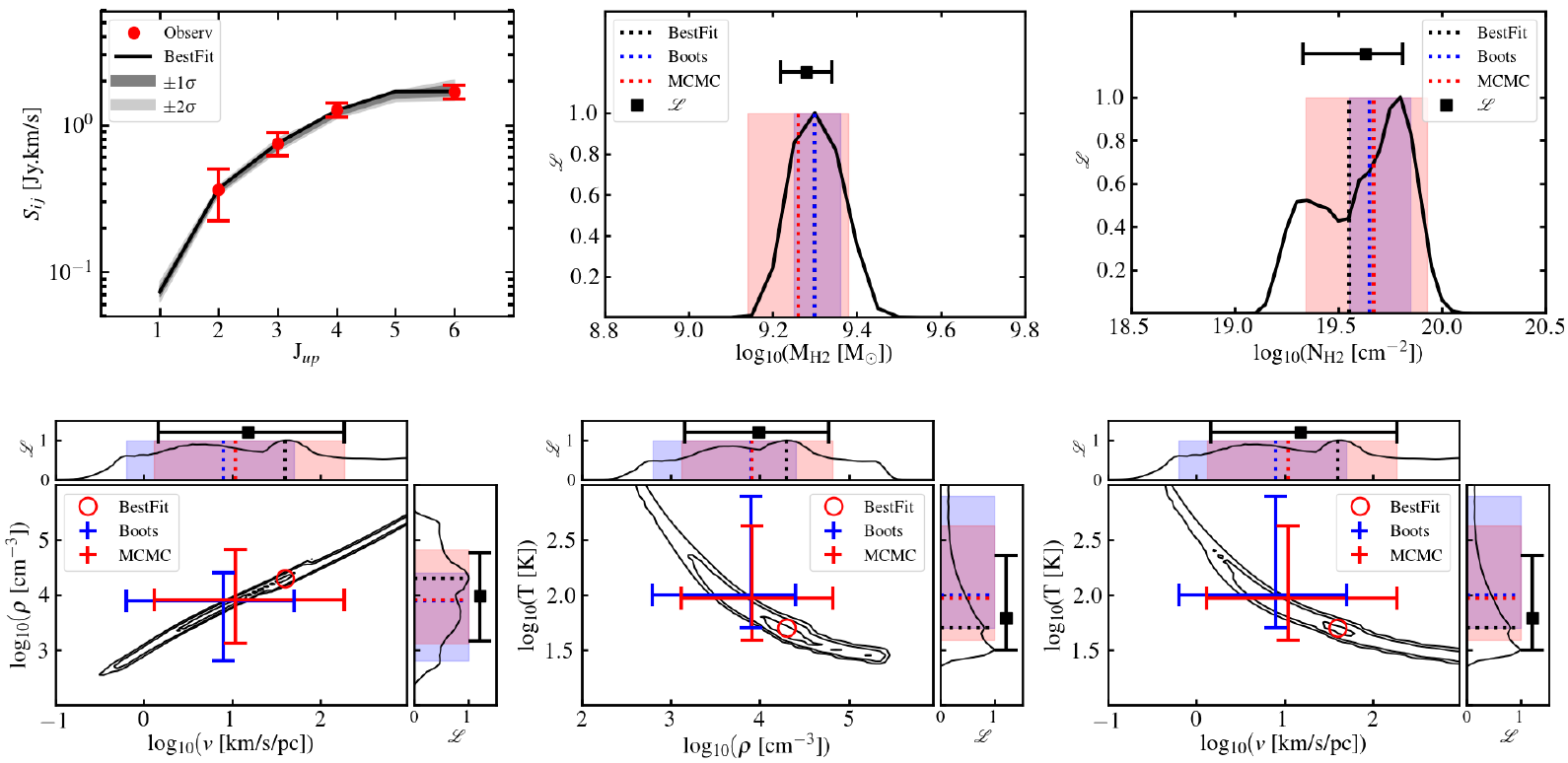}
		\caption{The same as Fig.~\ref{fig:lvgc1}, but for component\,{\sc II}.}
		\label{fig:lvgc2}
	\end{figure*}

	Figures~\ref{fig:lvgc1} and \ref{fig:lvgc2} show the results graphically for the analysis of CO and [CI] for components {\sc I} and {\sc II}, respectively, while Table~\ref{tab:lvg} presents the results quantitatively. In the figures, the top left panel shows the observed CO SLED as well as the best-fit (solid black line) and range of acceptable models (shaded grey regions). In the remaining plots, the central panel shows the surface likelihood distribution, with the best-fit value appearing as an open red-circle. Also, the MCMC (red) and Bootstrapping (blue) results appear as error-bars, where their centre indicates the 50$^{\rm th}$ percentile, and the extremes the 16$^{\rm th}$ and 84$^{\rm th}$ percentiles. These ranges are again displayed in the top and side panels as transparent regions (vertical dotted line indicating the 50$^{\rm th}$ percentile). Here, the likelihood distribution for each parameter is shown with a solid line, and its 16$^{\rm th}$, 50$^{\rm th}$, and 84$^{\rm th}$ percentiles are displayed with a black error-bar. In the side panels, the vertical black line indicates the best-fit value. For completeness, the predicted [CI]\,$^3P_1-^3P_0$ velocity integrated unmagnified fluxes for components {\sc I} and {\sc II} together is $0.72_{-0.09}^{+0.1}\,$Jy~km/s, which comprises the observed value of $0.77\pm0.09$.
	
	\begin{table}
		\caption{CO and [CI] SLED analysis in H$1429-0028$.}              
		\label{tab:lvg}      
		\centering                                      
		\begin{tabular}{crrrrr}          
			\hline\hline                        
			Pars.$^a$ & $\chi^2$ & Boots. & MCMC & ML & mean$^b$ \\    
			\hline                                   
			\multicolumn{6}{|c|}{Component {\sc I}}\\
			\hline
			log\,T & 1.60 & $1.6_{-0.3}^{+0.9}$ & $2.0_{-0.5}^{+0.6}$  & $1.8_{-0.4}^{+0.8}$  & $1.8_{-0.3}^{+0.4}$ \\
			log\,$n$ & 3.85 & $3.7_{-1}^{+0.9}$ & $3.4_{-0.6}^{+0.9}$  & $3.4_{-0.8}^{+0.9}$  & $3.5_{-0.5}^{+0.5}$ \\
			log\,$dv$ & 1.50 & $1_{-1}^{+1}$ & $1.0_{-0.6}^{+1}$  & $1.0_{-0.7}^{+1}$  & $1.0_{-0.4}^{+0.7}$ \\
			log\,N & 19.26 & $19.2_{-0.2}^{+0.4}$ & $19.2_{-0.2}^{+0.3}$  & $19.2_{-0.1}^{+0.2}$  & $19.2_{-0.1}^{+0.2}$ \\
			log\,$f$ & -0.94 & $-0.9_{-0.3}^{+0.3}$ & $-1.0_{-0.2}^{+0.2}$ & $-1.0_{-0.1}^{+0.2}$ & $-1.0_{-0.1}^{+0.1}$ \\
			log\,M & 9.27 & $9.3_{-0.1}^{+0.1}$ & $9.2_{-0.1}^{+0.1}$  & $9.21_{-0.06}^{+0.1}$  & $9.24_{-0.05}^{+0.06}$ \\
			\hline                                             
			\multicolumn{6}{|c|}{Component {\sc II}} \\
			\hline
			log\,T & 1.70 & $2.0_{-0.3}^{+0.9}$ & $2.0_{-0.4}^{+0.7}$  & $1.8_{-0.3}^{+0.6}$  & $1.9_{-0.2}^{+0.4}$ \\
			log\,$n$ & 4.30 & $3.9_{-1}^{+0.5}$ & $3.9_{-0.8}^{+0.9}$  & $4.0_{-0.8}^{+0.8}$  & $3.9_{-0.5}^{+0.4}$ \\
			log\,$dv$ & 1.60 & $0.9_{-1}^{+0.8}$ & $1.0_{-0.9}^{+1}$  & $1_{-1}^{+1}$  & $1.0_{-0.6}^{+0.6}$ \\
			log\,N & 19.55 & $19.7_{-0.1}^{+0.2}$ & $19.7_{-0.3}^{+0.3}$  & $19.6_{-0.3}^{+0.2}$  & $19.7_{-0.2}^{+0.1}$ \\
			log\,$f$ & -1.00 & $-1.1_{-0.2}^{+0.2}$ & $-1.2_{-0.2}^{+0.3}$ & $-1.1_{-0.2}^{+0.3}$ & $-1.1_{-0.1}^{+0.2}$ \\
			log\,M & 9.30 & $9.3_{-0.1}^{+0.1}$ & $9.3_{-0.1}^{+0.1}$  & $9.28_{-0.06}^{+0.06}$  & $9.28_{-0.05}^{+0.05}$ \\
			\hline
			\multicolumn{6}{l}{\footnotesize$^a$ The parameters in this column are:}\\
			\multicolumn{6}{l}{\footnotesize temperature, log\,T$\equiv$log(T~[K]);}\\
			\multicolumn{6}{l}{\footnotesize molecular gas density, log\,$n\equiv$\,log($n_{\rm H_2}$~[cm$^{-3}$]);}\\
			\multicolumn{6}{l}{\footnotesize velocity gradient, log\,$dv\equiv$\,log($dv~$[m/s/pc]);}\\
			\multicolumn{6}{l}{\footnotesize gas column density, log\,N$\equiv$log(N$_{\rm H_2}$~[cm$^{-2}$]);}\\
			\multicolumn{6}{l}{\footnotesize area filling factor, log\,$f\equiv~$log($<L^i_{obs}/L^i_{LVG}>_{weighted}$);}\\
			\multicolumn{6}{l}{\footnotesize and molecular gas mass, log\,M$\equiv$log(M$_{\rm H_2}$~[M$_\odot$]).}\\
			\multicolumn{6}{l}{\footnotesize$^b$ This column shows the log-mean between the ML, MCMC,}\\
			\multicolumn{6}{l}{\footnotesize and Bootstrap values.}\\
		\end{tabular}
	\end{table}
	
	\section{Discussion} \label{sec:disc}
	
	\subsection{Gas conditions} \label{sec:gascond}
	
	From Table~\ref{tab:lvg} and Figures~\ref{fig:lvgc1} and \ref{fig:lvgc2}, one can see that temperature, velocity gradient, and gas density are poorly constrained and correlated. As a result, there is no significant difference within the errors in these properties between the two velocity components. The molecular gas column density and mass are better constrained, and, although both velocity components have comparable gas masses (log(M$_{\rm H_2}$~[M$_\odot$])$\sim$9.26), the column density is $>1\sigma$ higher in {\sc II} (log(N$_{\rm H_2}$~[cm$^{-2}$])=$19.7_{-0.2}$ versus $19.2^{+0.2}$ in {\sc I}).
	
	Combined, components {\sc I} and {\sc II} have ${\rm M_{H_2}=3.6_{-0.6}^{+0.5}\times10^9 M_\odot}$ (where the errors assume the full range of uncertainty from all methods). For reference, this means that the H$_2$-to-dynamical-mass fraction (i.e., ${\rm M_{H_2} / M_{\rm DYN}}$) in the NS component is $8_{-6}^{+3}$\,per cent.
	
\begin{figure}
		\centering
		\includegraphics[width=0.5\textwidth]{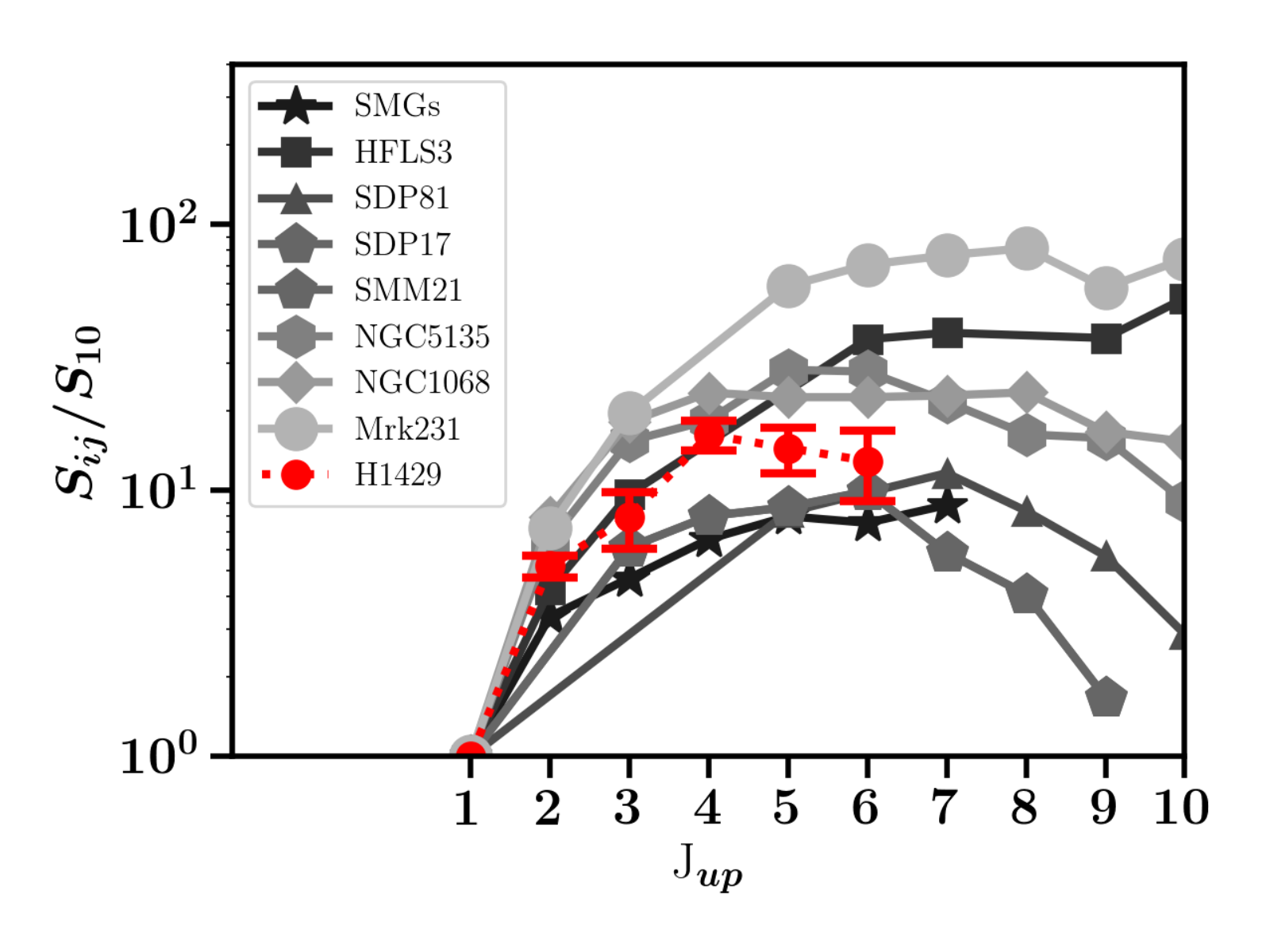}
		\caption{Comparing H1429-0028's CO SLED (the displayed values are total velocity-integrated ones) to other reported in the literature for: local IR-luminous galaxies \citep[NGC~5135, NGC~1068, Mrk~231;][]{rosenberg15}; SMGs \citep{bothwell13}; lensed SMGs \citep[SDP~81, SDP~17, SMM~J2135-0102, HFLS3;][]{danielson11,frayer11,harris12,lupu12,riechers13,alma15}. The CO\,(1-0) flux from H1429-0028 is that predicted from the analysis in Sec.~\ref{sec:lvg}.}
		\label{fig:sledlit}
	\end{figure}

	The CO SLED seems to be intermediate between those observed on average in SMGs and QSOs \citep[Fig.~\ref{fig:sledlit} and][]{carilli13,oteo17b}. Nevertheless, compared to the CO transitions luminosity-ratios presented by \citet[][table\,2 therein]{carilli13}, the resemblance to SMG-like ratios is strong: ${\rm L'_{32}/L'_{21}}=0.7\pm0.2$, ${\rm L'_{43}/L'_{21}}=0.78\pm0.09$, and ${\rm L'_{54}/L'_{21}}=0.45\pm0.08$ versus, respectively, average values of 0.78, 0.54, and 0.46 for SMGs or 0.98, 0.88, and 0.70 for QSOs \citep{carilli13}. Given the uncertainties and the current lack of evidence supporting the presence of an accreting super-massive black-hole \citep{timmons15,ma18}, H1429-0028 is regarded as a DSFG.
	
	Last, but not least, the gas conditions explaining the CO+[CI] SLED predict a CS\,(10-9) line flux $\sim$200 times lower than observed. This indicates that there is either a warm component inducing the population of higher J-levels in the CO-ladder, or the CS\,(10-9) emission is a result of differential magnification or chemical segregation in the background system. Follow-up observations of higher J-levels or deeper ALMA imaging can address this issue and the incidence of the high-velocity ($\sim\,$+500\,km/s) component.
	
	\subsection{Varying CO and Carbon abundances} \label{sec:cih2}
	
	\subsubsection{Changing $x_{\rm C}$ alone}
	
	The neutral Carbon abundance adopted in the LVG analysis in Section~\ref{sec:lvg} is $x_{\rm C}=10^{-4}$. This value is comparable to the abundance of $(8\pm4)\times10^{-5}$ that was found for a sample of submillimeter galaxies and quasar host galaxies at $z>2$ \citep{walter11}, and even in the centres of some local star-forming galaxies \citep{israelbaas03,jiao17}. However, in the literature, lower values are regularly adopted or reported, down to $x_{\rm C}\sim10^{-5}$ \citep{frerking89,sofia04,tomassetti14,israel15,jiao17}.
	
	As a first assessment, one can assume local thermal equilibrium (LTE) conditions and estimate the neutral carbon mass using the [CI](1-0) line alone (see Section~4.4 in M14 for details on estimating ${\rm M_{H_2}}$ based on [CI]). Considering a carbon excitation temperature of T$_{ex}=29\pm6$ \citep{walter11} and an observed [CI] luminosity from components {\sc I} and {\sc II} together of $L'_{\rm [CI]10}=(2.4\pm0.3)\times10^{9}~$K~km/s~pc$^2$, one retrieves ${\rm M_{[CI]}=(3.0\pm0.7)\times10^6 M_\odot}$. This implies a molecular gas mass of ${\rm M_{H_2}=(5\pm1)\times10^9 (x_{\rm C}/x^{alt}_{\rm C})~M_\odot}$, where $x_{\rm C}=10^{-4}$ and we have explicitly included the carbon abundance $x^{alt}_{\rm C}$ as an optional alternative to our adopted value. Within $1\sigma$, this estimated value agrees with the one obtained from the LVG analysis. One can see that the ${\rm M_{H_2}}$ estimate will vary linearly with the $x_{\rm C}/x^{alt}_{\rm C}$ ratio.
	
	We have also redone the LVG analysis assuming $x_{\rm CO}=10^{-4}$ and $x_{\rm C}=3\times10^{-5}$. This induces an increase in the ${\rm M_{H_2}}$ estimate by a similar factor to which the $x_{\rm C}$ was reduced by ($\sim3$), just as in the LTE assumption above. In this way, the molecular gas mass in components {\sc I} and {\sc II} would increase from ${\rm log(M_{H_2}~[M_\odot])} = 9.24$ to 9.75 and from 9.28 to 9.77, respectively. However, although the best-fit provides acceptable physical values, the peaks of the likelihood distributions and the MCMC posteriors imply very low volume densities $10^{2-3}\,$cm$^{-3}$ (even unconstrained in component {\sc I} within the model grid), and very low velocity gradients unconstrained down to the lowest values in the model grid ($dv=0.1{\rm~[km/s/pc]}$).
	
	\subsubsection{Assessing a metallicity dependence} \label{sec:xcz}
	
	Finally, one can assume that both $x_{\rm C}$ and $x_{\rm O}$ are proportional to metallicity (e.g., as in \citealt{wolfire10} or \citealt{narayanan12}), one arrives to the conclusion that $x_{\rm CO}\propto (Z')^n$, where $Z'=Z/Z_\odot$. The exponent $n$ here is assumed to be dependent on $x_{\rm C}$, $x_{\rm O}$, and the densities of these atoms, which already points to $n\sim2$ if one assumes that both C and O originate from the same process (stellar activity), and no CO dissociation occurs due to less shielding-effects at lower-metallicity.
	
	As a result, we have ran the LVG analysis assuming $x_{\rm C}=x^0_{\rm C}\times10^{(Z-Z_\odot)}$ and $x_{\rm CO}=x^0_{\rm CO}\times10^{2(Z-Z_\odot)}$. Here, $x^0_{\rm C}$ and $x^0_{\rm CO}$ are the neutral Carbon and CO abundances, respectively, adopted in Section~\ref{sec:lvg}, Z is the measured metallicity $12+log({\rm O/H})=8.49$ in $H1429-0028$ (see Section~\ref{sec:ametdep}) and Z$_\odot$ is the solar metallicity \citep[8.69;][]{asplund09}. This implies $x_{\rm C}=6.3\times10^{-5}$ and $x_{\rm CO}=4.0\times10^{-5}$. This approach results in total molecular mass of ${\rm M_{H_2}=5.6_{-0.9}^{+0.7}\times10^9 M_\odot}$, which is a factor of $1.6\pm0.3$ higher than the estimate reported in Section~\ref{sec:lvg}. Again, this factor is comparable to the one $x_{\rm C}$ was reduced by.
	
	\subsubsection{Accounting for the abundance uncertainty}
	
	As made clear in the previous sub-sections, the ${\rm M_{H_2}}$ estimate varies proportionally with the $x_{\rm C}$ assumption. Hence, as a conservative approach, and specifically considering the spread of the order of $\sim$40\,per cent in the abundances found in the sample studied by \citet{walter11}, a more realistic error estimate for the molecular gas estimate would be ${\rm M_{H_2}=4_{-2}^{+3}\times10^9 M_\odot}$, where the upper error estimate also accounts for the metallicity dependence uncertainty discussed in Section~\ref{sec:xcz}. This implies an H$_2$-to-dynamical-mass fraction in the NS component of $8_{-7}^{+6}$\,per cent.
	
	\subsection{CO-to-H$_2$ conversion-factor} \label{sec:co2h2}
	
	The CO/[CI] SLED analysis pursued in Sec.~\ref{sec:lvg} (Figs.~\ref{fig:lvgc1} and \ref{fig:lvgc2}) implies a velocity-integrated J=1-0 line-flux of $180_{-20}^{+20}~$mJy~km/s, implying a line luminosity of $L'_{\rm CO10}=1.0_{-0.1}^{+0.1}\times10^{10}~$K~km/s~pc$^2$ (assuming a magnification of $\mu_{\rm CO}\sim15$, Section~\ref{sec:diffmag}). This implies that the CO-to-H$_2$ conversion-factor in H$1429-0028$ is $\alpha_{\rm CO}={\rm M_{H_2}}/L'_{\rm CO10} = 0.4_{-0.2}^{+0.3}~$M$_\odot$/(K~km/s~pc$^2)$, assuming ${\rm M_{H_2}=4_{-2}^{+3}\times10^9 M_\odot}$.
	
	\subsubsection{Comparison with metallicity-dependent $\alpha_{\rm CO}$ relations} \label{sec:ametdep}
	
	As it has been proposed in the literature \citep[][for a review]{bolatto13}, the CO-to-H$_2$ conversion factor (${\rm M_{H_2+He}=\alpha_{CO} L'_{CO10}}$, ${\rm [\alpha_{CO}]=M_\odot/(K~km/s~pc^2)}$) appears to be dependent on gas-phase metallicity, where CO gradually becomes a poor tracer of H$_2$ with decreasing metallicity.
	
	Using 0.8-1.7\,$\mu$m grism data acquired by HST/WFC3, \citet{timmons15} assessed the gas-phase metalicity, 12+log(O/H), for H$1429-0028$. Their analysis revealed the A$+$B knot (Figure~1 in M14) as the brightest line emitter in the Einstein ring, resembling the knot flux-ratios seen at long-wavelengths ($\lambda\gtrsim1\,$mm), and in contrast with the rest-frame optical ones. This points to the fact that line emission is dominated by the North-South (NS) oriented component in the background system (Figure~8 in M14) also dominating the long-wavelength spectral range.
	
	Following the linear relation proposed by \citet{sobral15} between H$\alpha/$[NII] flux ratios and the logarithm of the H$\alpha+$[NII] equivalent width, one can use the former to estimate the metallicity based on the N2-index relations proposed by \citet{denicolo02} and \citet{pettini04}, giving, respectively, $12+log({\rm O/H})=8.7\pm0.2$ and $8.6\pm0.2$. Since the N2-index is known to saturate at metallicities close to solar, \citet{pettini04} proposed the O3N2-index which yields $8.5\pm0.2$. Assuming a solar metallicity of 8.69 \citep{asplund09}, H$1429-0028$ is observed as a system with a metallicity close to solar (e.g., O3N2 estimate is $1.3\sigma$ away).
	
	Following the metallicity-${\rm [\alpha_{CO}]}$ relation proposed by \citep[][which is based on the N2 calibration by \citealt{denicolo02}]{genzel12}, ${\rm \alpha_{CO}}$ is found to be 5.3~M$_\odot$/(K~km/s~pc$^2)$ (with an error factor of $\sim$1.7)\footnote{The reported error factor considers instrumental error alone, and not the scatter of the relation proposed by \citet{genzel12}.}. Based on the theoretical predictions by \citet{wolfire10} and \citet{glover11} \citep[which][show to best explain observations]{bolatto13}, and the empirical metallicity values just mentioned, the expected ${\rm \alpha_{CO}}$ is very much Milky-Way-like ($\sim4.3-7~$M$_\odot$/(K~km/s~pc$^2)$). An alternative theoretical approach by \citet{narayanan12} predicts a relation between ${\rm \alpha_{CO}}$ and both metallicity and CO line-intensity. This yields $\alpha_{\rm CO}=6\pm2~$M$_\odot$/(K~km/s~pc$^2)$, assuming the O3N2 metallicity index value and the $<W_{CO}>$ as measured in the source-plane CO(4-3) moment-0 map scaled to the CO(1-0) flux predicted from the SLED analysis.
	
	However, as briefly noted in M14, such high $\alpha_{\rm CO}$ values imply ${\rm M_{H_2}}$ values (${\rm \sim6-8\times10^{10}\,M_\odot}$) comparable or higher than the estimated dynamical mass (M$_{\rm DYN}=5_{-2}^{+3}\times10^{10} \,$M$_\odot$). As a result, these are deemed not realistic in the case of H$1429-0028$.
	
	\subsubsection{Comparison with the $L_{\nu_{\rm 850\mu m}}-L'_{\rm CO}$ relation}
	
	Since spectral line observations are time-consuming, the statistical approach proposed by \citet[][see also \citealt{hughes17,liang18}]{scoville16}, where the dust continuum is used as a tracer of molecular-gas, has become increasingly popular, since one can study numerous galaxies in a very inexpensive way. For completeness, the application of this relation to the case of H$1429-0028$ is reported here. We note however that the relation assumes a constant galactic-like ${\rm \alpha_{CO}}=6.5~$M$_\odot$/(K~km/s~pc$^2)$~\footnote{Note this value already includes a factor of 1.36 to account for elements heavier than Hydrogen.}, hence what should actually be compared here is CO\,(1-0) luminosity. Following Equation~16 in \citet[][Appendix~A.6]{scoville16}, we assume a power-law index of $3.9\pm0.4$ (M14) instead of 3.8, and $\alpha_{850}=(7\pm2)\times10^{19}\,$erg/(s~Hz~M$_\odot$). We also consider the {\sc mbb\_emcee}\footnote{https://github.com/aconley/mbb\_emcee} SED fit to $0.1-1.28\,$mm photometry in M14 to predict the observed flux at rest-frame 850\,$\mu$m, which avoids the need for the Rayleigh-Jeans approximation correction. This implies ${\rm M_{H_2}=2_{-1}^{+1}\times10^9 M_\odot}$ and $L'_{\rm CO10}=4_{-2}^{+2}\times10^{8}\,$K~km/s~pc$^2$. While the ${\rm M_{H_2}}$-mass estimate comprises the SLED analysis result, the CO(1-0) luminosity is a factor of $30_{-20}^{+10}$ lower than the predicted one (see values summarised in Section~\ref{sec:co2h2}). Nevertheless, H$1429-0028$ is a lensed galaxy and the errors are still relatively large. As a result, for the time being, this predicted-luminosity 1.8$\sigma$ difference ought to be considered as statistically insignificant.
	
	\subsection{The different ISM contents in the two background components} \label{sec:ismcont}
	
	To the depth of the current observations, it is observed that the gas and dust emissions are totally dominated by the NS component. Since the 1.28\,mm observations are probing the Rayleigh-Jeans tail of the dust continuum, we can use that information as a tracer of the ISM gas in the system \citep{scoville16}. Here, we chose to use this property to assess the relative ISM content between the EW and NS components. From Table~2 in M14, it is observed that the total 1.28\,mm flux is $6\pm1$\,mJy (all assigned to the NS component), while the observations reached an {\sc rms} of $78\,\mu$Jy, meaning a $3\sigma$ upper-limit of $0.23\,$mJy. Based on the latest lensing model, the magnification factors for the NS and EW components are $24\pm1$ and $11\pm1$, respectively~\footnote{The adopted magnification factor for this component is that measured at $K_s$-band, emission which is dominated by the EW component.}. As a result, the EW gas and dust content is $<9^{+9}$\,per cent \footnote{The uncertainty here takes into account a factor of 2 in uncertainty to account for dust-to-gas ratio variations.} of that observed in the NS component (i.e., ${\rm M_{H_2} [M_\odot]}<3^{+4}\times10^8$). A more direct comparison requires deeper 1\,mm dust-continuum data or higher spatial resolution observations of low-J transitions (J$_{\rm up}\leq3$), where a less excited CO SLED from the EW component is expected to peak, and NIR spectral observations to assess the dynamic properties of the EW component.
	
	\subsection{Comparison with SED fitting analysis} \label{sec:cavunc}
	
	In M14, we adopted a two step approach to retrieve the background emission uncontaminated by the foreground one. We used \textit{galfit} in the high spatial resolution \textit{F110W}, $H$, and $K_s$ bands imaging to deblend foreground and background emission. Based on this, we later used \textit{MagPhys} \citep[adopting the default low-$z$ template library;][]{dacunha08} to determine the foreground emission in the remainder low spatial resolution imaging. On top of this, we further corrected the background rest-frame UV/optical emission for foreground obscuration (adopting different scenarios and taking them into account in the photometry error budget) and differential magnification. Nevertheless, \citet{ma18} has recently shown that, even after this considered approach, depending on the star-formation history assumed for the template library used for the SED fitting, one can retrieve significantly different conclusions. Also, \citet{zhang18} also shows that the initial mass function in sources alike H1429 may be top-heavy, which was not considered in M14 nor \citet{ma18}. Although these issues affect even unlensed systems, H1429 has also been shown to be comprised by two spatially separated components, one dominating the UV/optical spectral range, the other the FIR-radio one. This goes against the underlying assumption of both MagPhys and CIGALE \citep{noll09} where the stellar and dust components are co-spatial.
	
	As a result, in this manuscript, we avoid any discussion involving conclusions based on the SED fitting done in previous works. Since the molecular and dust content is mostly dominated by the NS component, this work is essentially a description of its properties based on the mm observations reported here and in M14.
	
	\section{Conclusions} \label{sec:conc}
	
	In this paper, the gravitationally-lensed galaxy merger HATLAS\,J142935.3-002836 (H1429-0028) at $z=1.027$ presented in \citet{messias14} is characterised in further detail. Specifically, recent APEX observations with SHeFI-APEX2 and the recent SEPIA-Band~5 instrument targeting, respectively, the CO transitions J=6-5 and 3-2, allowed us to assess the ISM gas content in the background system.
	
	Thanks to the recent APEX observations and previous ALMA ones, a continuous coverage of the CO-SLED from J$_{\rm up}$=2 to 6 is now available, together with the [CI]\,$^3P_1-^3P_0$ transition. We have identified three different velocity components comprising the spectra with velocity centroids at $v_c^{I}=-130_{-6}^{+6}$\,km/s, $v_c^{II}=131_{-5}^{+5}$\,km/s, and $v_c^{III}=500_{-30}^{+20}$\,km/s (Sec.~\ref{sec:linedec}). It is observed that they contribute differently to each transition, but, based on the updated lensing model we find that the two brightest components are equally magnified by a factor of $\sim15$, while the fainter one is magnified by $\sim12$. We also show that the high-velocity component is morphologically much smaller than the others, which seems to agree with its expected higher degree of excitation. Only the two main components {\sc I} and {\sc II} were considered in the analysis, since the high velocity one is unconstrained in the [CI] spectrum. See Sec.~\ref{sec:lines}, \ref{sec:linedec}, and \ref{sec:diffmag}. 
	
	Assuming a large velocity gradient scenario and a combined statistical approach (Maximum Likelihood, Markov Chain Monte Carlo, Bootstrap), the molecular gas content in H$1429-0028$ is estimated to be ${\rm M_{H_2}=4_{-2}^{+3}\times10^9 M_\odot}$, where the error accounts for the uncertainty in neutral Carbon and CO abundances. This amount of gas comprises about $8_{-7}^{+6}$\,per cent of the dynamical mass in the NS component. As a result, at the time of observation, this star-formation event is expected to turn only up to 15\,per cent ($1\sigma$ upper-limit) of the total (dynamical) mass into stars. No major excitation differences between components\,{\sc I} and {\sc II} are observed, but the column density is apparently higher toward component\,{\sc I}. Averaging over the many statistical approaches and over the two main velocity components, the gas temperature, volume density, and velocity-gradient are constrained within a factor of 3 on a galaxy-wide view. These parameters are estimated to be T$\sim$70\,K, $\log({\rm n [cm^{-3}]})\sim3.7$, and $dv\sim10$\,km/s/pc. The gas column density is constrained within a factor of 1.4 to be $\log({\rm N [cm^{-2}]})\sim19.4$. See Sections~\ref{sec:lvg} and \ref{sec:cih2}.
	
	Compared to galaxy samples in the literature, H$1429-0028$ is observed to have a DSFG-like CO-SLED \citep[in line with the lack of evidence thus far supporting the presence of AGN;][]{timmons15,ma18} and, based on the predicted CO~(1-0) velocity-integrated flux, a CO-to-H$_2$ conversion factor ($\alpha_{\rm CO}=0.4_{-0.2}^{+0.3} ~$M$_\odot$/(K~km/s~pc$^2)$). See Section~\ref{sec:gascond}.
	
	The spatially-resolved dust continuum map allows us to have a first assessment of the relative ISM gas content between the two background components. We estimate that the EW component is very gas and dust poor with a content less than $9^{+9}$\,per cent of what is observed toward the NS component (i.e., ${\rm M^{EW}_{H_2}}<3^{+4}\times10^8$~M$_\odot$). See Section~\ref{sec:ismcont}.
	
	\section*{Acknowledgements}
	
	HM thanks the opportunity given by the ALMA Partnership to work at the Joint ALMA Observatory via its Fellowship programme. HM acknowledges support by FCT via the post-doctoral fellowship SFRH/BPD/97986/2013.
	
	We thank insightful discussion with Paola Di Matteo from the GalMer team while attempting to understand this system by the use of galaxy-merger models.
	
	We thank the comments provided by Asantha Cooray.
	
	NN acknowledges support from Conicyt (PIA ACT172033, Fondecyt 1171506, and BASAL AFB-170002).
	
	ZYZ and IO acknowledges support from the European Research Council in the form of the Advanced Investigator Programme, 321302, {\sc cosmicism}.
	
	SD is supported by the UK STFC Rutherford Fellowship scheme.
	
	E.I.\ acknowledges partial support from FONDECYT through grant N$^\circ$\,1171710.
	
	D.R. acknowledges support from the National Science Foundation under grant number AST-1614213.
	
	MJM acknowledges the support of the National Science Centre, Poland through the POLONEZ grant 2015/19/P/ST9/04010; this project has received funding from the European Union's Horizon 2020 research and innovation programme under the Marie Sk{\l}odowska-Curie grant agreement No. 665778.
	
	Based on data products from observations made with APEX telescope under programmes IDs C-087.F- 0015B-2011, 087.A-0820, 088.A- 1004, and 097.A-0995.
	
	This paper makes use of the following ALMA data: ADS/JAO.ALMA\#2011.0.00476.S. ALMA is a partnership of ESO (representing its member states), NSF (USA) and NINS (Japan), together with NRC (Canada), MOST and ASIAA (Taiwan), and KASI (Republic of Korea), in cooperation with the Republic of Chile. The Joint ALMA Observatory is operated by ESO, AUI/NRAO and NAOJ.
	
	This research made use of iPython \citep{perezgranger07}, Numpy \citep{vanderwalt11}, Matplotlib \citep{hunter07}, SciPy \citep{jones01}, and Astropy \citep[a community-developed core Python package for Astronomy,][]{astropy13}.

	\bsp	
	\label{lastpage}
\end{document}